\documentclass[referee]{aa} 
\usepackage{natbib}
\usepackage{graphicx}
\usepackage{txfonts}
\begin{document}
   \title{A new view of quiet-Sun topology from Hinode/SOT}

   \author{S. R\'egnier
          \and
          C. E. Parnell
	  \and
	  A. L. Haynes
          }

   \offprints{S. R\'egnier}

   \institute{School of Mathematics and Statistics, University of St Andrews,
   St Andrews, Fife, KY16 9SS, Scotland, UK\\
              \email{stephane@mcs.st-andrews.ac.uk}
             }

   \date{Received ; accepted }

 
  \abstract
   {With the recent launch of the Hinode satellite our view of the nature and
   evolution of quiet-Sun regions has been improved. In light of the
   new high resolution observations, we revisit the study of the quiet
   Sun's topological nature.}
   {Topology is a tool to explain the complexity of the magnetic
   field, the occurrence of reconnection processes, and the heating of the
   corona. This Letter aims to give new insights to these different topics.}
   {Using a high-resolution Hinode/SOT observation of the line-of-sight
   magnetic field on the photosphere, we calculate the three dimensional
   magnetic field in the region above assuming a potential field. From the 3D
   field, we determine the existence of null points in the magnetic
   configuration.}
   {From this model of a continuous field, we find that the distribution of
   null points with height is significantly different from that reported in
   previous studies. In particular, the null points are mainly located above
   the bottom boundary layer in the photosphere (54\%) and in the chromosphere
   (44\%) with only a few null points in the corona (2\%). The density of null
   points (expressed as the ratio of the number of null points to the number of
   photospheric magnetic fragments) in the solar atmosphere is estimated to be
   between 3\% and 8\% depending on the method used to identify the number of
   magnetic fragments in the observed photosphere.}
   {This study reveals that the heating of the corona by magnetic reconnection
   at coronal null points is unlikely. Our findings do not rule out the heating
   of the corona at other topological features. We also report the topological
   complexity of the chromosphere as strongly suggested by recent observations
   from Hinode/SOT.  }


   \maketitle

\section{Introduction}

The quiet-Sun magnetic field, the so-called magnetic carpet \citep{tit98}, is
thought to be divided into two main parts: closed field lines and open field
lines \citep[see e.g.][]{gab76}. The closed field lines are low-lying
intertwined field lines responsible for the complexity of the field. In
particular, \citet{clo03} revealed that 50\% of the magnetic flux in the
quiet-Sun atmosphere is closed down within 2.5 Mm above the solar surface, with
almost 5--10\% extending to 25 Mm above the solar surface. It is
assumed that this complexity, in addition to the fast recycling of the magnetic
flux \citep{hag03, clo05}, is responsible for nano- and microflares that
contribute to coronal heating \citep{par02}. Thus, understanding the complex
topology of the quiet-Sun magnetic field is key to understanding the
coronal heating problem. 

As known from decades of observations, the quiet-Sun magnetic field can be
depicted by a large number of positive and negative sources. The source motions
are characterized by emergence, coalescence, fragmentation and cancellation of
magnetic polarities. These motions increase the complexity of the field. In
order to analyse the complexity of the magnetic field, we study the existence
and distribution of null points above an observed region of the quiet Sun. Even
though null points are not the only relevant topological element in a magnetic
configuration, they are a good proxy for the complexity of the field and for
the existence of magnetic reconnection. Previous studies \citep{schr02, lon03,
clo04a} have tackled this topic with different approaches. \citet[][hereafter
ST02]{schr02} modelled the quiet-Sun photosphere by point charges randomly
distributed with a balanced magnetic flux. They also compared these with TRACE
EUV images \citep{han99} and SOHO/MDI line-of-sight magnetograms \citep{sch95}.
The coronal magnetic field was then defined analytically under the potential
field assumption (i.e. no current present in the configuration) as detailed in
\citet{lon96}. These authors found that in the coronal volume there was
approximately one null point per charge with 9\% of the null points located in
the corona (indicating that 91\% of null points are photospheric null points). 
In the point charge approach used by ST02 the bottom boundary layer has a zero
vertical component of the magnetic field everywhere except at each point
charge. This assumption leads to many null points at the base. Their comparison
with TRACE images showed that there is no link between the EUV brightenings and
the number of null points, except for one location in their example.
\citet[][hereafter LBP03]{lon03} extended this study  by producing a theory to
determine the density of null points in random potential magnetic fields
anchored in the photosphere. They tested their theory using a point charge
approach and permitted an imbalance of flux thus producing open field lines.
The authors found similar results to ST02 for the density of null points in the
corona (above the photosphere). In addition LBP03 showed that the density of
null points depends on how the sources are distributed in the photosphere and
on the degree of imbalance of flux in the region. It is briefly mentioned in
LBP03 that most of the photospheric null points are an artefact of the method
used (no vertical field outside sources) and are proportional to the number of
sources used in the realisation. In addition to simulated quiet-Sun data,
\citet[][hereafter CPP04]{clo04a} used a SOHO/MDI photospheric magnetogram to
study the complexity of the field. They converted the observed magnetogram into
an array of point charges using a clumping approach to define the location and
flux of the sources. In a large field-of-view of
240$\arcsec\times$240$\arcsec$, they found only 286 sources and a null point
density of 0.038 in the corona (above the boundary).

In this Letter, we derive the null point density from a potential field
extrapolation of a photospheric magnetogram obtained by Hinode/SOT. However,
instead of assuming a point charge model, we consider a continuous magnetic field
through the base. We compare our results with the statistical modeling of the
quiet Sun by ST02, LBP03 and CPP04.

\section{Observation and model}
\label{sec:obs_mod}

\subsection{Hinode/SOT data}

We use a line-of-sight magnetogram derived from the observed Stokes parameters
I and V by the NFI (Narrowband Filter Imager) of the Hinode/SOT (Solar Optical
Telescope, \citeauthor{tsu08} \citeyear{tsu08}). The observation was recorded
on June 24, 2007 at 22:09 UT,  and views a region slightly off the disc center.
The NaI D1 resonance line at 5896 \AA~is used. Although the core of the line
($< \pm$70 m\AA) comes from chromospheric radiation \citep[see e.g.,][]{bru92},
the NFI magnetogram is considered as a photospheric magnetogram since the
measurements are made in the wings of the line at a spectral resolution of 90
m\AA.  The field-of-view is 141\arcsec$\times$161\arcsec~(102 Mm$\times$116 Mm)
with a pixel width of 0.16\arcsec~(880$\times$1012 pixels). The pixel width of
SOT/NFI is 3.8 times better than that for high resolution observations taken by
SOHO/MDI. The analysis, reduction and calibration of this dataset are detailed
in \citet{par08}.

\begin{figure}[!ht]
\centering
\includegraphics[width=.9\linewidth]{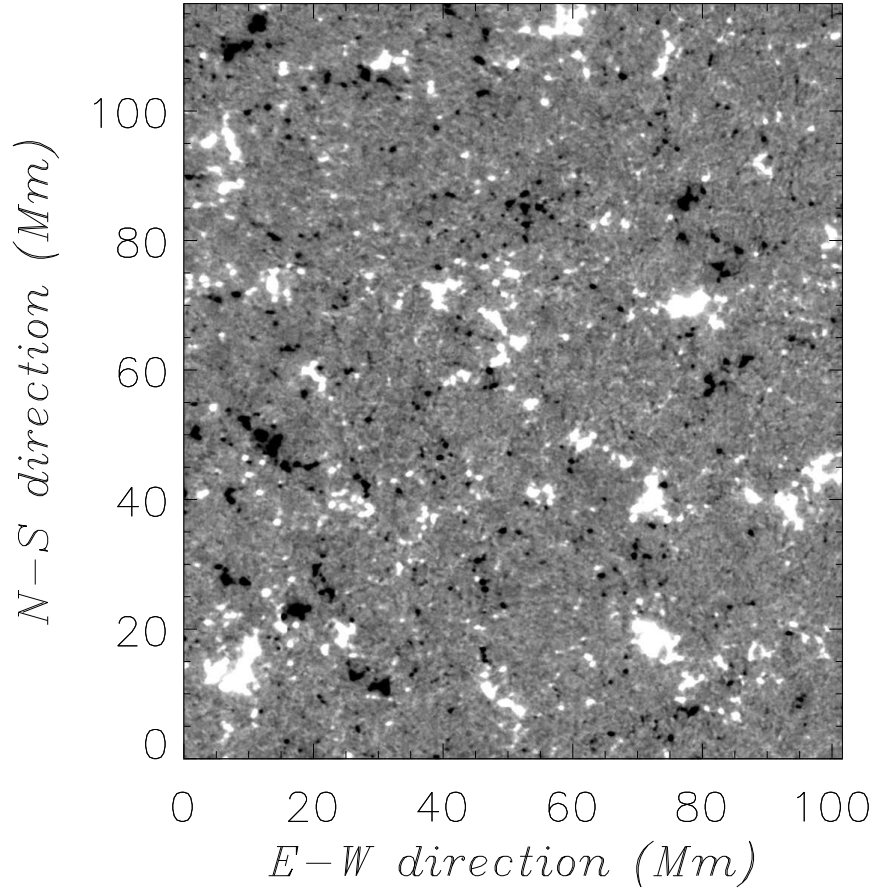}
\caption{Distribution of the line-of-sight magnetic field on the photosphere
observed by Hinode/SOT NFI on June 24, 2007 at 22:09 UT. The image is saturated
at $\pm$50 G. Positive (negative) polarities are white (black). North is up.}
\label{fig:bz_hinode}
\end{figure}

In Fig.~\ref{fig:bz_hinode}, we plot the distribution of the line-of-sight
magnetic field in a quiet-Sun region as observed by SOT/NFI. The maximum 
absolute field strength is 717 G. The net flux is 3.58 $\times$ 10$^{20}$ Mx
and the total unsigned flux is 1.5 $\times$ 10$^{21}$ Mx. The imbalance of
magnetic flux is then $\Delta \phi = +24\%$. The sign indicates that there is
an excess of positive magnetic flux.

\subsection{Potential field model}

Assuming that the coronal magnetic field is essentially in equilibrium,
neglecting the gas pressure and gravity forces, as well as the currents flowing
along field lines, we obtain the potential field assumption satisfying $\vec
\nabla \wedge \vec B = \vec 0$. The potential field corresponds to the minimum
energy state. Potential field extrapolation methods require only the vertical
(resp. radial) magnetic field component in Cartesian geometry (resp. spherical
geometry) at the bottom boundary layer \citep[e.g.][]{sch64, sem68}. This is a
well-posed boundary value problem giving a unique solution for prescribed
boundary conditions on all boundaries. In the past, several types of field
distributions have been used as the base boundary condition: (i) the point
charge method which permits an analytical expression for the coronal field to
be derived assuming that each magnetic fragment is reduced to a single point
\citep[e.g.,][]{lon96}, (ii) the discrete source method for which the
polarities have a finite width but there is no field outside, (iii) the
continuous field method where the observed magnetic field is discretised and
then extrapolated numerically within a finite box. We here apply the last
method using the Hinode/SOT data detailed in the previous Section as a bottom
boundary condition (defined on a smaller grid of 292$\times$336 pixels due to
computational limitations). As the magnetic flux is not balanced, we impose
open boundary conditions on the sides and top of the box allowing field lines
to enter or leave the box. Due to the assumed high complexity of the quiet-Sun
field near the photosphere, we use a stretched grid in the vertical direction
with a finer grid near the photosphere.  

\subsection{Null point finder}

A classical method to find null points in a field configuration is based on the
computation of the Poincar\'e index \citep{gre92}. Unfortunately the method
fails for weakly and strongly nonlinear fields \citep{hay07}. Recently, a more
reliable and stable method for linear and moderately nonlinear fields has been
developed: the trilinear method \citep{hay07}. By comparison to
\citeauthor{gre92}'s method, the trilinear method is more appropriate for
numerical fields whilst \citeauthor{gre92}'s method is only really suitable for
analytical fields where any grid can be easily made finer if the method fails.
The trilinear method has been successfully applied to different numerical
experiments \citep{hay07}. In order to apply this method, the considered vector
field has to be linear or moderately nonlinear within a cell. Here we have had
to use a stretched grid in the $z$ direction with smaller grid spacing near the
photosphere to ensure that the complexity of the field is well resolved in the
low atmosphere.


\begin{figure}[!ht]
\centering
\includegraphics[width=.9\linewidth]{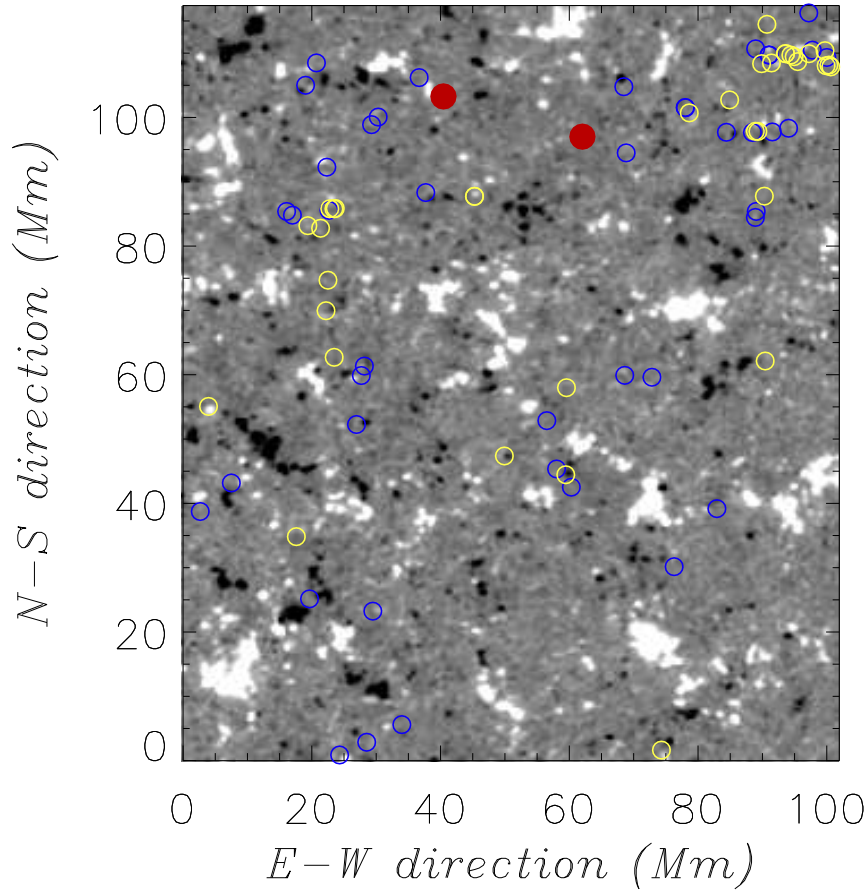}
\caption{Location of the null points in the photosphere (blue circles), in the
chromosphere (yellow circles) and in the corona (filled red circles).
Background image: line-of-sight SOT magnetogram used for the potential field
extrapolation (defined on a smaller grid of 292$\times$336 pixels due to
computational limitations) saturated at $\pm$50 G.}
\label{fig:qs0_null}
\end{figure}

	
\section{Discussion and conclusions}
\label{sec:disc}

We extrapolate the coronal field under the potential field assumption from the
photospheric line-of-sight component provided by Hinode/SOT data
(Fig.~\ref{fig:bz_hinode}). Then we look for null points in the three
dimensional magnetic configuration using the trilinear method.

In Fig.~\ref{fig:qs0_null}, we plot the spatial distribution of the null points
in the {\em xy}-plane (parallel to the photospheric plane). We find 80 null
points distributed in height from the bottom boundary to about 5 Mm above
which  no null points are found. None of the null points lie exactly on the
bottom boundary. The spatial distribution in the {\em xy}-plane is seemingly
random, in particular there exist areas as large as a network cell (20
Mm$\times$20 Mm) without null points. Few null points lie above concentrations
of flux. 

\begin{figure}
\centering
\includegraphics[width=.41\linewidth]{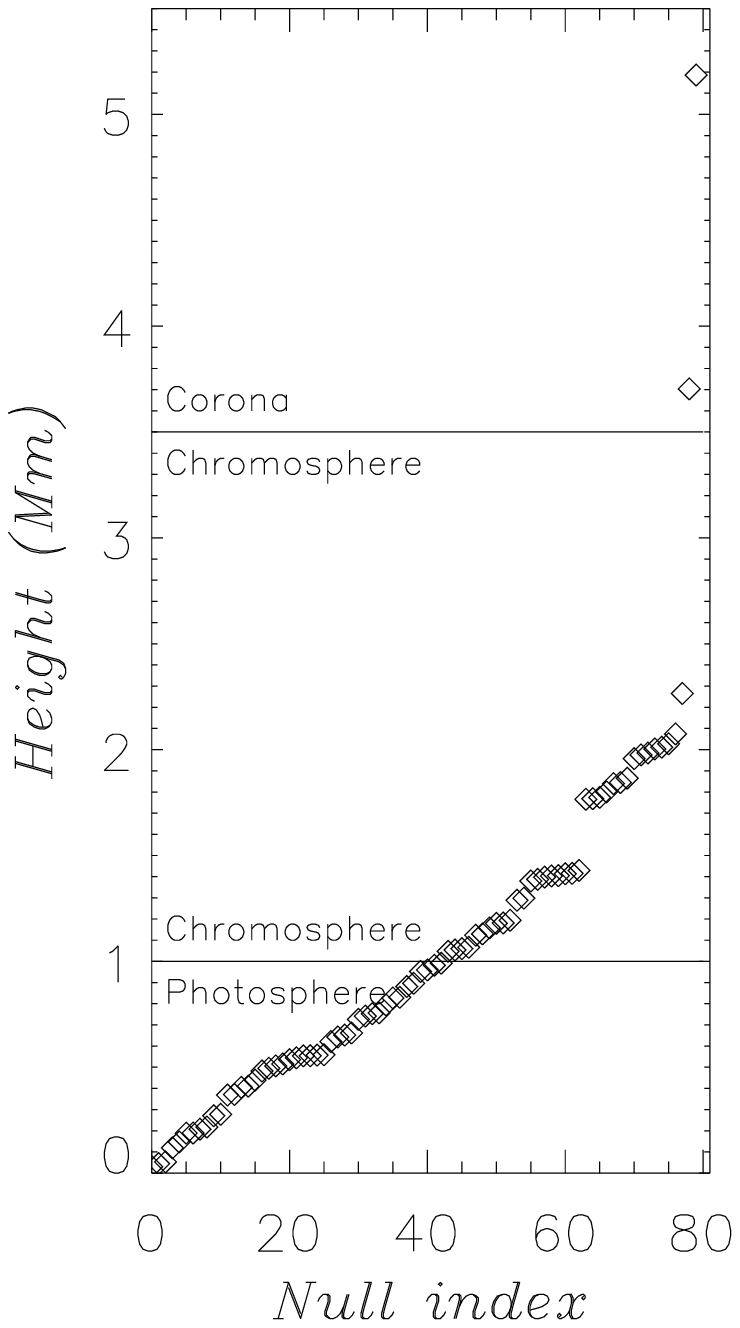}
\includegraphics[width=.58\linewidth]{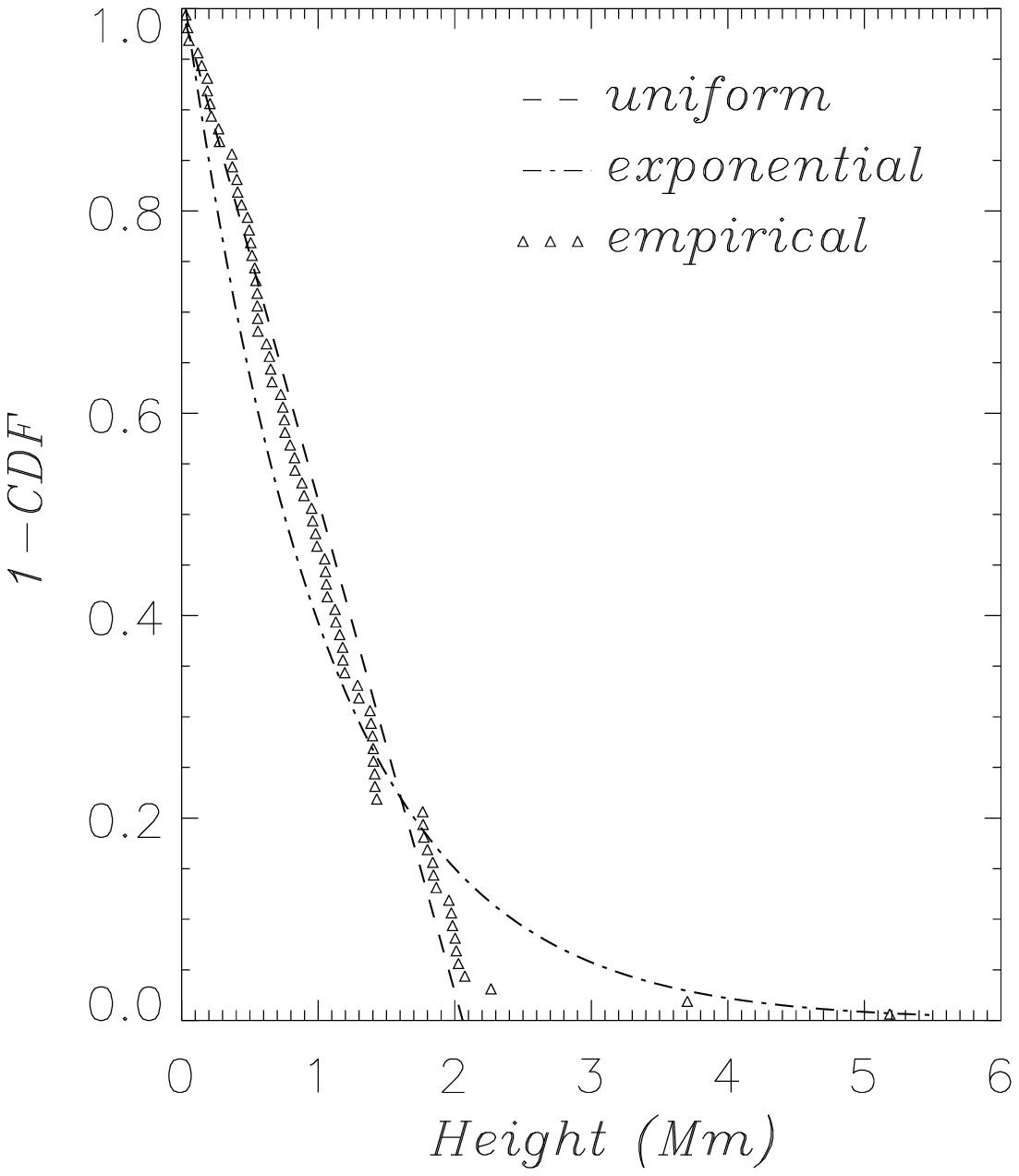}
\caption{Left: Distribution of null points with height (diamonds). 43 null
points are in the photosphere below 1 Mm, 35 null points in chromosphere
between 1 and 3.5 Mm, and 2 null points are in the corona. Right: Survival
functions (1-CDF, Cumulative Distribution Function) for the distribution of null
point heights (triangles), the exponential fit for the atmospheric null points
(dot-dashed curve) and the uniform fit for the photospheric and chromospheric
null points (dashed curve).}
\label{fig:qs0_h}
\end{figure}

We now assume a typical solar atmosphere:  the photosphere from 0 to 1 Mm, the
chromosphere from 1 to 3.5 Mm, and then the corona. This atmosphere is depicted
in Fig.~\ref{fig:qs0_h} left by the horizontal solid lines. We assume that the
height of formation of the NaI D1 line provided by SOT/NFI is at the base of
photosphere. Under these assumptions, we find:
\begin{itemize}
\item[-]{43 null points (54\%) in the photosphere;}
\item[-]{35 null points (44\%) in the chromosphere;}
\item[-]{2 null points (2\%) in the corona.}
\end{itemize}

As shown in Fig.~\ref{fig:qs0_h} right, there is a uniform distribution of null
points below 2 Mm (approximately 70-95\% of the characteristic source
separation) whilst the distribution of all atmospheric null points is well
fitted by an exponential distribution with a characteristic length of 1.04 Mm
which is about half of the source separation as in ST02.

We define the density of null points as the ratio of the number of the null
points  to magnetic fragments in order to compare our results with ST02, LBP03
and CPP04. To derive this quantity, we need to determine the number of magnetic
fragments in the quiet-Sun region. There are several methods to identify the
fragments as reviewed by \cite{def07}. For comparison, we only use two methods:
the clumping approach and the peak approach. Following \citet{par08}, there are
more than one thousand contiguous magnetic flux fragments assuming a threshold
of 18 G in the observed quiet-Sun region using the clumping approach, giving a
density of null points less than 7.5\% (1063 fragments). The peak approach with
a threshold of 18 G gives 2680 fragments leading to a density of null points of
3\%.  

The results obtained here from the extrapolation of a continuous photospheric
distribution may be compared to those found using the point charge method of
simulated data used by  ST02, LBP03 and CPP04 (not reported here). The
estimated densities of null points are summarised in Table~\ref{tab:dens}. With
the point charge method used by ST02, the density of null points on the
photosphere (bottom boundary layer) is about one null point per charge, and 9\%
of null points in the corona (above the bottom boundary). In LBP03, it is not
possible to derive the density of null points on the bottom layer due to the
definition of the random distribution of point charge sources although there
will be many, but the density in the corona is between 7--9\% Clearly the
density of null points estimated using the point charge approach and the
estimate made using magnetic fragments determined from the clumping
identification method in a continuous field can be reconciled by simply
removing the base boundary layer from consideration. If this is done, the
density of null points in the solar atmosphere (photosphere + chromosphere +
corona) is about 7--9\%. Nevertheless if the number of sources is derived from
the peak method, then the density of null points is much less (3\%), which is
comparable to the density obtained with observed magnetogram by CPP04. Note
that compared to our study with the peak approach, CPP04 used a field-of-view
five times larger but involving 10 times fewer sources. Clearly, the comparison
of the clumping results from this work and CPP04s results suggests that the
spatial density of null points may also vary with resolution. This will be
investigated in detail in a subsequent paper. This suggests that the density of
null points is not an appropriate quantity to describe the complexity of an
observed quiet-Sun region as it relies on a magnetic field model and a
definition of magnetic fragments.


\begin{table}[!h]
\caption{Summary of the null point density depending on the location in the
solar atmosphere for the different methods (see text for detail)}
\begin{center}
\begin{tabular}{l c c c c c}
\hline
\hline\\[-0.3cm]
density & ST02 & LBP03 & CPP04 & \multicolumn{2}{c}{this work} \\
\hline
photosphere & & & clump & clump & peak \\
\qquad {\it boundary} & 1.001 & N/A & 1.05 & 0 & 0 \\
\qquad {\it above} & N/A & N/A & N/A & 0.04 & 0.016 \\
chromosphere & N/A & N/A & N/A & 0.033 & 0.013 \\
corona & 0.093 & 0.07--0.09 & 0.038 & 0.002 & 0.0007 \\
total atmos. & 0.093 & 0.07--0.09 & 0.038 & 0.075 & 0.03\\
\hline
grand total & 1.094 & 0.07--0.09 & 1.088 & 0.075 & 0.03 \\
\end{tabular}
\end{center}
\label{tab:dens}
\end{table}


As pointed out by ST02, the topology of the quiet Sun is important in
understanding the possible heating mechanisms of the corona. By comparing EUV
images and photospheric magnetograms, ST02 showed that there is no correlation
between the coronal EUV brightenings (from the FeXII line at 195\AA) and the
location of null points, with the exception of one point. From our study, we
see that only 2 null points are in the corona and are potentially observable in
such a hot corona. Most of the null points should be observable in the
chromosphere. The complexity of the chromosphere was recently highlighted by
the CaII images from Hinode/SOT \citep[see e.g.,][]{dep07}. Due to the low
spatial density of null points in the corona, the heating of the corona by
magnetic reconnection at null points located in the corona can then be
excluded. Nevertheless reconnection can occur at different locations with
(e.g., separators, separatrix surfaces) or without (e.g., hyperbolic flux
tubes) null points.  

In a forthcoming paper, we will develop the study of the topology of the
quiet-Sun field by looking at the connectivity of field lines, the stability
and nature of null points and their time evolution. This study is important to
understand the dynamic behaviour of heating in a complex topology. 

\begin{acknowledgements}
We thank the referee for his/her comments which helped to improve this
Letter. We thank the UK STFC for financial support (STFC RG). CEP is funded by
STFC on an Advanced Fellowship. Hinode is a Japanese mission developed and
launched by ISAS/JAXA, with NAOJ as domestic partner and NASA and STFC (UK) as
international partners. It is operated by these agencies in co-operation with
ESA and NSC (Norway). The computations have been done using the XTRAPOL code
developed by T. Amari (Ecole Polytechnique, France). We also acknowledge the
financial support by the European Commission through the SOLAIRE network
(MTRN-CT-2006-035484).
\end{acknowledgements}



\end{document}